\documentclass[letterpaper, 10 pt, conference]{ieeeconf}  

\IEEEoverridecommandlockouts                              

\usepackage{graphics} 
\usepackage{epsfig} 
\usepackage{mathptmx} 
\usepackage{times} 
\usepackage{amsmath} 
\usepackage{amssymb}  
\usepackage{url}
\usepackage{multirow}
\title{\LARGE \bf
Enhancing Urban Sensing Utility with Sensor-enabled Vehicles and Easily Accessible Data
}

\author{Hui Zhong$^{1,2}$, Qing-Long Lu$^{3}$, Qiming Zhang$^{1}$, Hongliang Lu$^{1,2}$, Xinhu Zheng$^{1,2}$
\thanks{Corresponding to Xinhu Zheng. {\tt\small xinhuzheng@hkust-gz.edu.cn}}
\thanks{$^{1}$Hui Zhong, Qiming Zhang, Hongliang Lu and Xinhu Zheng are with INTR thrust, System Hub, the Hong Kong University of Science and Technology (Guangzhou), Guangdong, 511400, China.
        }%
\thanks{$^{2}$Hui Zhong, Hongliang Lu and Xinhu Zheng are also with Guangdong Provincial Key Lab of Integrated Communication, Sensing and Computation for Ubiquitous Internet of Things.}
\thanks{$^{3}$Qiong-Long Lu is with the Department of Civil and Environmental Engineering, National University of Singapore, Singapore 117576, Singapore.
        }%
}

\begin{document}

\maketitle
\thispagestyle{empty}
\pagestyle{empty}

\begin{abstract}
Urban sensing is essential for the development of smart cities, enabling monitoring, computing, and decision-making for urban management.
Thanks to the advent of vehicle technologies, modern vehicles are transforming from solely mobility tools to valuable sensors for urban data collection, and hold the potential of improving traffic congestion, transport sustainability, and infrastructure inspection.
Vehicle-based sensing is increasingly recognized as a promising technology due to its flexibility, cost-effectiveness, and extensive spatiotemporal coverage. 
However, optimizing sensing strategies to balance spatial and temporal coverage, minimize redundancy, and address budget constraints remains a key challenge. 
This study proposes an adaptive framework for enhancing the sensing utility of sensor-equipped vehicles. 
By integrating heterogeneous open-source data, the framework leverages spatiotemporal weighting to optimize vehicle selection and sensing coverage across various urban contexts. 
An entropy-based vehicle selection strategy, \texttt{Improved OptiFleet}, is developed to maximize sensing utility while minimizing redundancy. 
The framework is validated using real-world air quality data from 320 sensor-equipped vehicles operating in Guangzhou, China, over two months. 
Key findings show that the proposed method outperforms baseline strategies, providing up to 5\% higher sensing utility with reduced fleet sizes, and also highlights the critical role of dynamic urban data in optimizing mobile sensing strategies.
\end{abstract}

\section{Introduction}

Urban sensing is one of the cornerstones of smart city initiatives, supporting systematic monitoring, analysis, and decision-making for urban management and governance applications \cite{ghahramani2020urban}. 
Traditionally, vehicles are regarded merely as vessels for personal mobility, seamlessly transporting individuals across urban landscapes. 
However, the rapid proliferation of vehicle stocks has introduced a series of traffic-related challenges, including traffic congestion, air pollution, parking shortages, and rising energy demands, all of which increasingly threaten the livability and sustainability of urban areas \cite{wang2023aggravated, kirschner2020parking}. 
As these urban pressures mount, vehicles are progressively perceived not merely as enablers of mobility, but as \textit{troublemakers} contributing to urban issues \cite{plante2025investing}. 

In parallel with the evolution of smart cities, the advent of ubiquitous sensing technologies has fostered the emergence of an innovative data collection paradigm known as mobile crowdsensing \cite{suhag2023comprehensive}. 
Among emerging approaches, vehicle-based sensing, also known as drive-by sensing (DS), has attracted considerable attention due to its low deployment cost, high flexibility, and broad spatiotemporal coverage \cite{ji2023survey}. 
As a result, modern vehicles (usually equipped with various sensors) are increasingly viewed as potential \textit{problem solvers}, evolving from simple mobility tools into distributed sensing nodes that monitor detailed urban metrics at the street level, often outperforming traditional fixed monitoring systems \cite{cheng2022integrated}.
Modern vehicles play an ambivalent role in current urban transportation systems, acting both as contributors to congestion and environmental challenges, and as potential platforms for innovative solutions. 
How to effectively harness their benefits while mitigating their negative impacts remains a compelling and enduring challenge for the transportation community.

Despite its promise, building an effective spatiotemporal sensing network with sensors-enabled vehicles remains a critical challenge, mainly due to divergent requirements in different application scenarios \cite{han2024exploring}. 
Air quality monitoring, for example, requires a fine-grained and continuous spatial coverage, while infrastructure health inspections often require localized observations \cite{di2022vehicular, bonola2016opportunistic}. 
To address the challenges of optimizing urban sensing with sensor-enabled vehicles under diverse application requirements, this study makes the following key contributions:
\begin{itemize}
    \item Proposes an adaptive framework for vehicle-based urban sensing that maximizes sensing utility under budget constraints using a maximum entropy-based optimization approach.
    \item Introduces a spatiotemporal weighting mechanism into the entropy computation to incorporate task-relevant urban context, enabling the framework to flexibly adapt to various sensing scenarios and seamlessly accommodate additional sensing tasks. 
    Specifically, our method fuses heterogeneous, publicly available datasets, including static street-view imagery and dynamic traffic congestion indices, among others, to characterize the urban context. 
    The contextual features extracted through deep learning are then used to construct adaptive spatiotemporal sensing weights, facilitating task-specific optimization of mobile sensing performance across diverse application scenarios. 
    \item Further develops a marginal gain-based vehicle selection strategy, \texttt{Improved OptiFleet}, under budget constraints to maximize sensing utility via greedy optimization.
    \item Validates the proposed framework through case studies using large-scale real-world air quality data collected from 320 sensor-enabled vehicles operating in Guangzhou, China, between March 1 and April 30, 2023.
\end{itemize}



The remainder of the paper is organized as follows. Section~\ref{sec:literature} reviews related work. Section~\ref{sec:method} presents the proposed methods for sensing utility quantification and vehicle fleet optimization. Section~\ref{sec:results} discusses the experimental results. Finally, Section~\ref{sec:conclusions} concludes the paper and outlines directions for future research.

\section{Related work}\label{sec:literature}
Urban sensing utility is fundamentally determined by the spatiotemporal sensing patterns of the DS. Addressing this problem requires solving two key challenges: (\romannumeral 1) developing rigorous methods to quantify spatiotemporal sensing utility, and (\romannumeral 2) designing effective vehicle selection strategies under resource and budget constraints. 

For quantifying sensing utility, existing studies have primarily adopted two classes of indicators: (\romannumeral 1) the number of visits (or samples) and the number of distinct taxis sensing a location, and (\romannumeral 2) the gap interval between successive sensing events within a given subregion \cite{tonekaboni2020spatio}. The first type quantifies sensing frequency, while the second measures the temporal continuity of observations \cite{o2019quantifying}. Typically, a location is considered adequately sensed once a predefined number of samples, denoted as $K$, has been collected. This threshold-based approach defines a subregion as `sensed' when it receives more than $K$ samples. Recent studies \cite{hou2025advancing} show that setting excessively high revisit frequency requirements can significantly reduce DS coverage, highlighting a trade-off between sensing granularity and spatial coverage. However, most studies assume consistent spatiotemporal sensing requirements and overlook the varying needs of different application scenarios \cite{ji2023survey}.

Building on the established sensing metrics, numerous studies have developed vehicle selection strategies that seek to maximize coverage or minimize deployment costs, typically employing greedy optimization methods \cite{ji2023survey, di2022vehicular}. 
For example, Zhao et al. \cite{zhao2015opportunistic} introduced the \textit{inter-cover time metric (ICT)} to quantify the frequency with which a grid cell is opportunistically covered by vehicles. Based on historical mobility traces, they proposed a selection algorithm to minimize the number of vehicles required to achieve a predefined coverage ratio. While this method prioritizes coverage utility, it often leads to spatiotemporal redundancy and lacks adaptability to varying spatiotemporal requirements across different application scenarios. 
Khan et al.\cite{khan2016autonomous} proposed an information-centric framework, InfoRank, which autonomously ranks location-aware information to optimize sensing performance under budget constraints. However, their approach was designed for vehicular ad hoc networks (VANETs) and evaluated solely through simulations using the ndnSIM module in NS-3, without validation on real-world data.
Tonekaboni et al.\cite{tonekaboni2020spatio} proposed the Utility-Aware Redundancy Minimization Algorithm, which selects vehicles based on the number of grid cells they traverse. To improve the representation of spatiotemporal variability, the method incorporates population and variation in the sensing target to adjust the selection priority. Although this adjustment enhances spatiotemporal balance, the approach depends on prior knowledge of the sensing target and has limited adaptability across different application contexts.  In contrast, Fan et al. \cite{fan2021towards} proposed deploying dedicated vehicles to enhance coverage uniformity, though this significantly increases operational costs and limits scalability.

Overall, existing vehicle fleet selection strategies are highly sensitive to the definition of sensing utility \cite{han2024exploring, mora2019towards}. When utility functions inadequately represent individual vehicle contributions, selection results may be inefficient or redundant. Many studies rely primarily on spatial coverage as a proxy for sensing utility, which can lead to excessive information redundancy and neglect the contextual relevance of sensed data.
Furthermore, most prior approaches lack validation using real-world sensing outputs, limiting their applicability in practical deployments. Defining a sensing utility function that effectively balances spatiotemporal informativeness and resource constraints remains a key challenge.

\section{Methodology}\label{sec:method}
To enhance urban sensing utility, this section presents a structured methodology comprising three key steps: quantifying sensing utility across spatiotemporal domains, estimating adaptive spatiotemporal weights using open-source urban data, and optimizing vehicle fleet selection under budget constraints.
\subsection{Sensing utility quantification}

We begin by discretizing the sensing domain into a set of spatial grid cells $\mathcal{G}$ and temporal intervals $\mathcal{T}$. 
Specifically, the study area is divided into $G$ spatial cells at a certain resolution, such as 500 m $\times$ 500 m, and the time horizon is partitioned into $T$ time intervals over a day. 
Together, $\mathcal{G} \times \mathcal{T}$ defines the complete spatiotemporal sensing space. 
For each candidate vehicle $v$, we define $p_{g,t}(v)$ as the normalized frequency (or probability) that vehicle $v$ visits spatial cell $g \in \mathcal{G}$ during time interval $t \in \mathcal{T}$. 
Each vehicle $v$ is associated with a sensing deployment cost $c_v$, and the total available budget is denoted by $B$. 
Denote the set of vehicle candidates by $\mathcal{V}$. 
Given a selected fleet $S \subseteq \mathcal{V}$, the probability that at least one vehicle in $S$ covers a specific spatiotemporal point $(g,t)$ is computed as
\begin{equation}
P_{g,t}(S) = 1 - \prod_{v \in S} \left(1 - p_{g,t}(v)\right).
\label{eq:coverage}
\end{equation}
Then, the sensing utility of fleet $S$, denoted by $f(S)$, is defined as the cumulative expected coverage across all $(g,t)$ cells, which is given by 
\begin{equation}
f(S) = \sum_{g \in \mathcal{G}} \sum_{t \in \mathcal{T}} w_{g,t}P_{g,t}(S).
\label{eq:utility}
\end{equation}
where $w_{g,t}$ denotes the weight of spatiotemporal point $(g,t)$ within the sensing space, representing the importance of the associated cell.
This sensing utility function offers two merits: (\romannumeral 1) The probability-based coverage indicator illustrated in Equation~\eqref{eq:coverage} is more robust compared to the binary one and supports the application of the maximum entropy theory to vehicle fleet optimization that is expounded on in Section~\ref{sec:veh_opt}; 
(\romannumeral 2) Employing $w_{g,t}$ to account for the importance heterogeneity across the sensing space enhances the flexibility of the function in adapting to various urban sensing scenarios.

\subsection{Spatiotemporal weights estimation}

Clearly, the spatiotemporal weights play a critical role in the accuracy of the sensing utility function with respect to any given urban sensing problem.
In this study, we propose to utilize easily accessible data to achieve reliable estimations of the weights.
Specifically, static street-view imagery and dynamic traffic congestion indices, retrieved via the Baidu Map API\footnote{\url{https://map.baidu.com/}}, are adopted to capture the spatiotemporal contextual features. 
These datasets have been widely demonstrated to effectively reflect local variations in pollutant concentrations \cite{zhong2025predicting}. 

\begin{figure}[!htbp]
    \centering
    \includegraphics[width=1\linewidth]{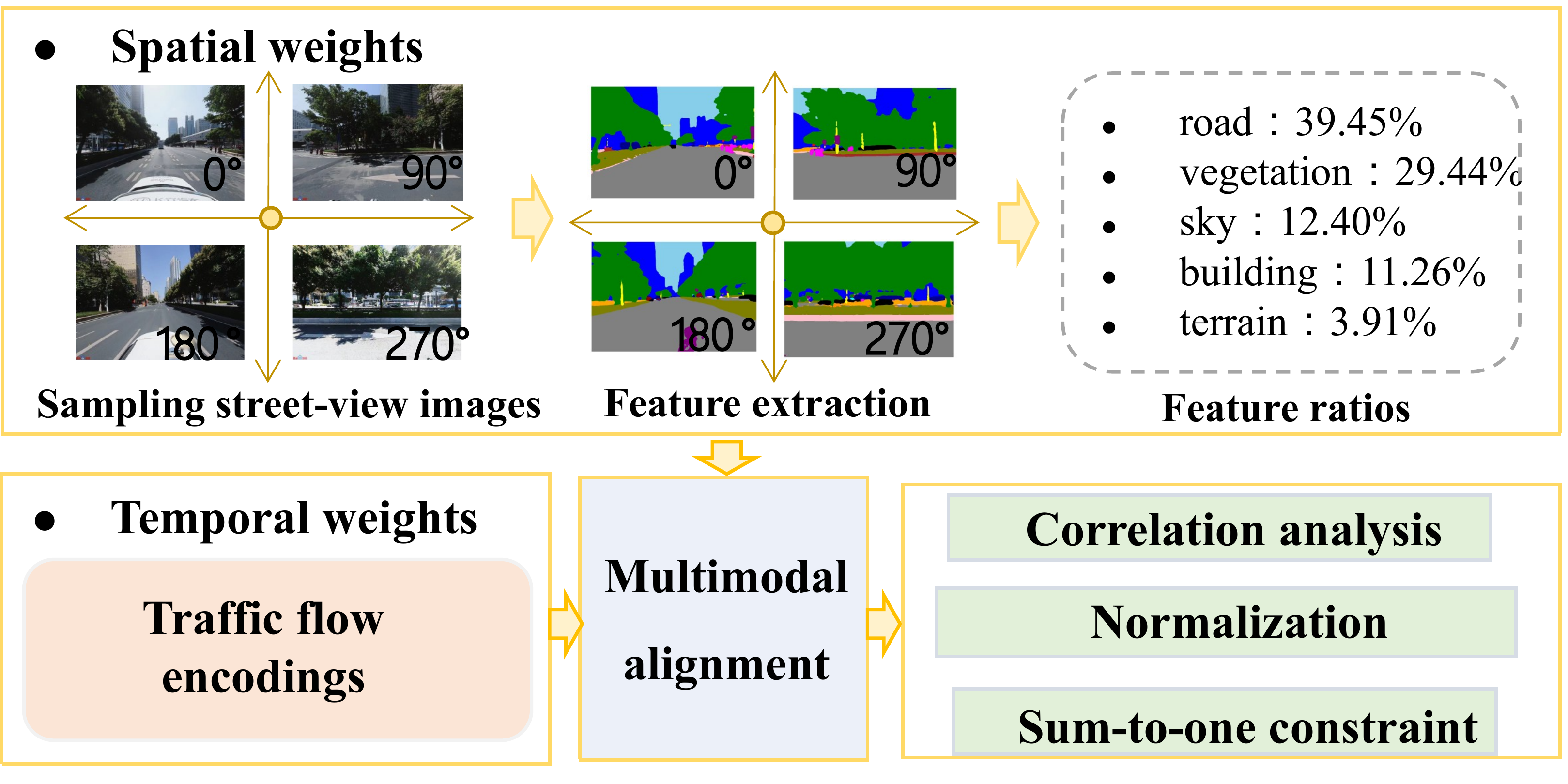}
    \caption{Framework for calculating adaptive spatiotemporal sensing weights based on Mask2Former algorithm.}
    \label{fig1}
\end{figure}

The overall procedure for acquiring, processing, and utilizing open datasets to construct spatial weights is illustrated in Fig.~\ref{fig1}. 
First, low-quality images were filtered out using OpenCV-based image processing techniques \cite{garcia2017multi}, considering factors such as overexposure, underexposure, blurring, and color channel distortion. 
Next, street scene features were extracted from the filtered images using the Mask2Former scene parsing algorithm \cite{cheng2022masked}. 
Correlation analysis was then conducted to quantify the relationships between the extracted features and pollutant concentrations, resulting in a set of correlation coefficients for each feature-pollutant pair. 
These coefficients were subsequently normalized to generate the final spatial weighting factors. 

\subsection{Vehicle fleet optimization}\label{sec:veh_opt}

Our objective is to select a subset of vehicles $S \subseteq \mathcal{V}$ that maximizes the overall sensing utility under the budget constraint.
Mathematically, the problem can be described as
\begin{equation}
\begin{aligned}
    \max_{S\subseteq \mathcal{V}} \quad & f(S) \\
    \text{s.t.} \quad & \sum_{v \in S} c_v \leq B
\end{aligned}
\end{equation}

Assuming that individual vehicle visits are independent, then $f(S)$ is monotone and submodular since $p_{g,t}(v)$ lies within $[0,1]$. 
Accordingly, the problem reduces to a classical Maximum Coverage Location Problem (MCLP), for which a simple greedy algorithm can guarantee a $(1 - 1/e)$-approximation to the optimal solution \cite{church1974maximal}.

Given the NP-hard nature of MCLP, finding an exact solution is computationally intractable for large-scale instances. 
Hence, we adopt the classic Greedy-Add heuristic to address this problem \cite{ali2017coverage}. 
To maximize the sensing utility under budget constraints, we develop \texttt{OptiFleet}, a marginal gain-based selection strategy. 
It iteratively computes the marginal utility gain of adding a new vehicle $u$ to the current selected subset $S$, prioritizing vehicles that contribute the most to overall spatiotemporal coverage per unit cost. 
Specifically, for each spatiotemporal point $(g, t)$, the updated coverage probability after including $u$ is calculated as follows:
\begin{equation}
P_{g,t}(S \cup \{u\}) = 1 - \left(1 - P_{g,t}(S)\right) \left(1 - p_{g,t}(u)\right)
\end{equation}
Accordingly, the marginal gain $\Delta_u$ of adding vehicle $u$ to the current set $S$ can be expressed as:
\begin{equation}
\Delta_u = \sum_{(g, t) \in \mathcal{G} \times \mathcal{T}} w_{g, t} \left( P_{g,t}(S \cup \{u\}) - P_{g,t}(S) \right)
\end{equation}

At each step, the candidate vehicle with the maximum marginal utility gain per unit cost is selected, until the total budget $B$ is exhausted. The selection rule is formalized as:
\begin{equation}
u^* = \arg\max_{u \in V \setminus S} \frac{1}{c_u} \Delta_u
\end{equation}

\begin{table*}[t]
\caption{Comparison of sensing utility across different selection strategies}
\centering
\small
\begin{tabular}{cccccccc}
\hline
\multirow{2}{*}{Algorithm} & \multicolumn{7}{c}{Size of vehicle fleets (Numbers)}                                                                                                          \\ \cline{2-8} 
                           & 10                  & 50                   & 100                  & 150                  & 200                  & 250                  & 300                  \\ \hline
TSUB                         & 4.23                & 18.17                & 33.26                & 46.93                & 58.47                & 68.69                & 78.13                \\
RA                         & 3.88                & 17.56                & 31.92                & 45.44                & 57.54                & 68.35                & 77.94                \\
\texttt{OptiFleet}                         & {\textbf{4.41}} & 18.97                & 34.03                & 47.41                & 59.14                & 68.90                & 78.46                \\
\texttt{Improved OptiFleet}                & 4.17                & { \textbf{19.00}} & { \textbf{35.15}} & { \textbf{48.94}} & { \textbf{60.91}} & {\textbf{70.93}} & {\textbf{79.26}} \\ \hline
\end{tabular}

\vspace{0.5em}
\footnotesize\textit{Note*: The best performance in each case is highlighted in boldface.}
\label{tab:comparison_sensing}
\end{table*}

\subsection{Improved OptiFleet}
Although \texttt{OptiFleet} effectively improves overall coverage, it tends to favor vehicles that repeatedly visit already well-covered areas, especially in dense urban centers. 
To mitigate this limitation, we incorporate a complementary information-theoretic perspective based on \textit{Shannon entropy}, which prioritizes the diversity of spatiotemporal coverage rather than its frequency. 
Intuitively, vehicles that explore a broader range of spatiotemporal points provide greater informational value to the sensing system. 
The entropy of vehicle $v$'s trajectory is defined as:
\begin{equation}
    H(v) = - \sum_{(g, t) \in \mathcal{G} \times \mathcal{T}} p_{g,t}(v) \log_2 \left( p_{g,t}(v) \right)
\end{equation}

When selecting vehicles sequentially, it is important to account for redundant information captured by previously selected vehicles. 
To address this, we define the effective coverage of candidate vehicle $u$ given the current subset $S$ as:
\begin{equation}
    \tilde{p}_{g,t}(u|S) = p_{g,t}(u)  \left(1 - C_{g,t}(S)\right)  w_{g,t}
\end{equation}
This adjusted probability downweights spatiotemporal points that are already well-covered. 
Based on this, the effective entropy of vehicle $u$ is calculated as:
\begin{equation}
    H(u|S) = - \sum_{(g, t) \in \mathcal{G} \times \mathcal{T}} \tilde{p}_{g,t}(u|S) \log_2 \tilde{p}_{g,t}(u|S)
\end{equation}
This entropy-based metric serves as a proxy for the new spatiotemporal information contributed by vehicle $u$ to the selected fleet $S$, providing a complementary criterion to marginal coverage. 
Vehicles can then be selected based on:
\begin{equation}
    u^* = \arg\max_{v \in V \setminus S} \frac{1}{c_u} H_u(S)
\end{equation}

\section{Results}\label{sec:results}

\subsection{Study area and data preprocessing}
We conducted experiments over a two-month period (March 1 to April 30, 2023) across the major administrative districts. The study area is the political, economic, and cultural core of Guangzhou \cite{zhong2023dynamic}. It covers approximately 279.63 km\textsuperscript{2} and exhibits high levels of human activity, with a permanent population density of about 21,812 inhabitants per square kilometer \cite{gzstats2023}. To facilitate spatial analysis, the area was divided into regular grid cells of 500 m × 500 m, resulting in a total of 3,811 cells.

Taxi trajectory and air quality data were obtained from the sensor-enabled vehicles, provided by the Guangzhou Taxi Company. The dataset includes GPS records collected at approximately 15-second intervals, capturing the longitude, latitude, and timestamp of each sensor-equipped taxi. In parallel, four key atmospheric pollutants, nitric oxide (NO), nitrogen dioxide (NO\textsubscript{2}), fine particulate matter (PM\textsubscript{2.5}), and inhalable particulate matter (PM\textsubscript{10}), were continuously recorded during the data collection period. 
We applied a Hidden Markov Model (HMM) map-matching algorithm to the taxi trajectories and the road network extracted from OpenStreetMap\footnote{\url{https://www.openstreetmap.org/}} to accurately align GPS points to the network \cite{newson2009hidden}.

\subsection{Model performance}


To compare the performance of the models, this study introduces three baseline models. The first baseline, referred to as the Taxi-based Sensing Utility Baseline (TSUB), is based on the sensing utility definition with the number of distinct taxis \cite{han2024exploring}. 
The number of distinct vehicles whose trajectories intersect with grid cell $g$ during time period $t$ is denoted by $N_{g,t}$. The corresponding sensing utility is defined as $\xi(N_{g,t}) = \sum_{n=1}^{N_{g,t}} 1/n^{\alpha} = (N_{g,t})^{\beta}$, where $\beta$ is determined based on the minimum and maximum spatial weights and visiting times, following the approach in \cite{han2024exploring}. 
In this study, $\beta$ was estimated to be 1.85.
The second baseline, referred to as the \texttt{OptiFleet} algorithm, selects vehicles through an optimization strategy that aims to maximize sensing utility. The third baseline is the random assignment (RA), where vehicles are selected randomly without any optimization. 
As a result, the methods differ not only in their selection strategies but also in their definitions of coverage. 
To evaluate the performance of different methods, vehicle fleets are first selected according to their respective strategies. 
Their sensing utilities are then calculated using the coverage function (\ref{eq:utility}) proposed in this study. 
Table\ref{tab:comparison_sensing} compares the coverage results under consistent experimental settings.

Overall, the proposed method outperforms the baseline algorithms in terms of sensing utility. Under limited fleet size and tight budget constraints, the TSUB strategy achieves the best performance by maximizing spatial coverage with minimal resources. However, as fleet size increases, the proposed entropy-driven method (\texttt{Improved OptiFleet}), consistently delivers superior performance by enhancing spatiotemporal diversity and reducing redundancy in collected data. Whereas, the advantage of the proposed method gradually diminishes as the fleet size becomes larger. Specifically, under the same fleet size, our method improves sensing accuracy by approximately 5\% compared to the worst-performing method (RA).

\begin{figure}[htbp]
  \centering
  \includegraphics[width=\linewidth]{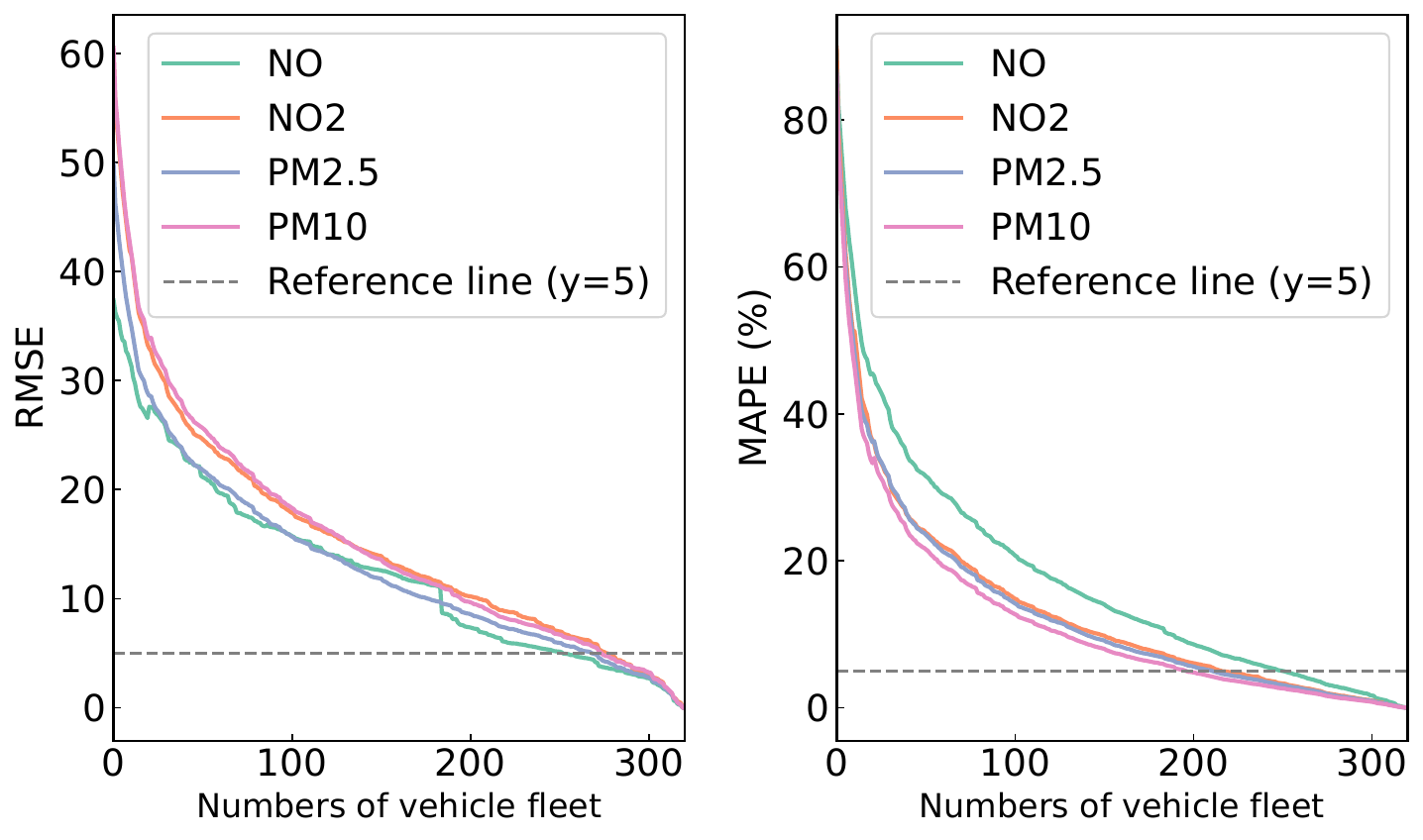}
  \caption{Estimation errors of the selected vehicle fleets with the increase in fleet size.}
  \label{fig2}
\end{figure}

To assess the performance of the \texttt{OptiFleet} algorithm, this study employs two evaluation metrics: RMSE and MAPE, which measure the estimation errors between fleets of different sizes and the full fleet, with the full fleet's measurements considered the ground truth. The comparison results are shown in Fig.~\ref{fig2}. Although sensing accuracy exhibits some variability across different pollutants, a consistent overall pattern is observed: the sensing utility increases with fleet size, but the rate of improvement gradually diminishes. This behavior is consistent with Gossen’s First Law, which states that the marginal utility of a resource declines as its consumption increases\cite{moss1984laws}. Specifically, as additional vehicles are incorporated into the fleet, their marginal contribution to novel spatiotemporal information progressively diminishes as a result of increased redundancy and overlapping coverage. Notably, when the selected fleet size reaches approximately 200 vehicles, the average MAPE across space and time drops below 5\%. As the fleet size increases to 250 vehicles, the RMSE further decreases to below 5.

\begin{figure}[!htbp]
  \centering
  \includegraphics[width=\linewidth]{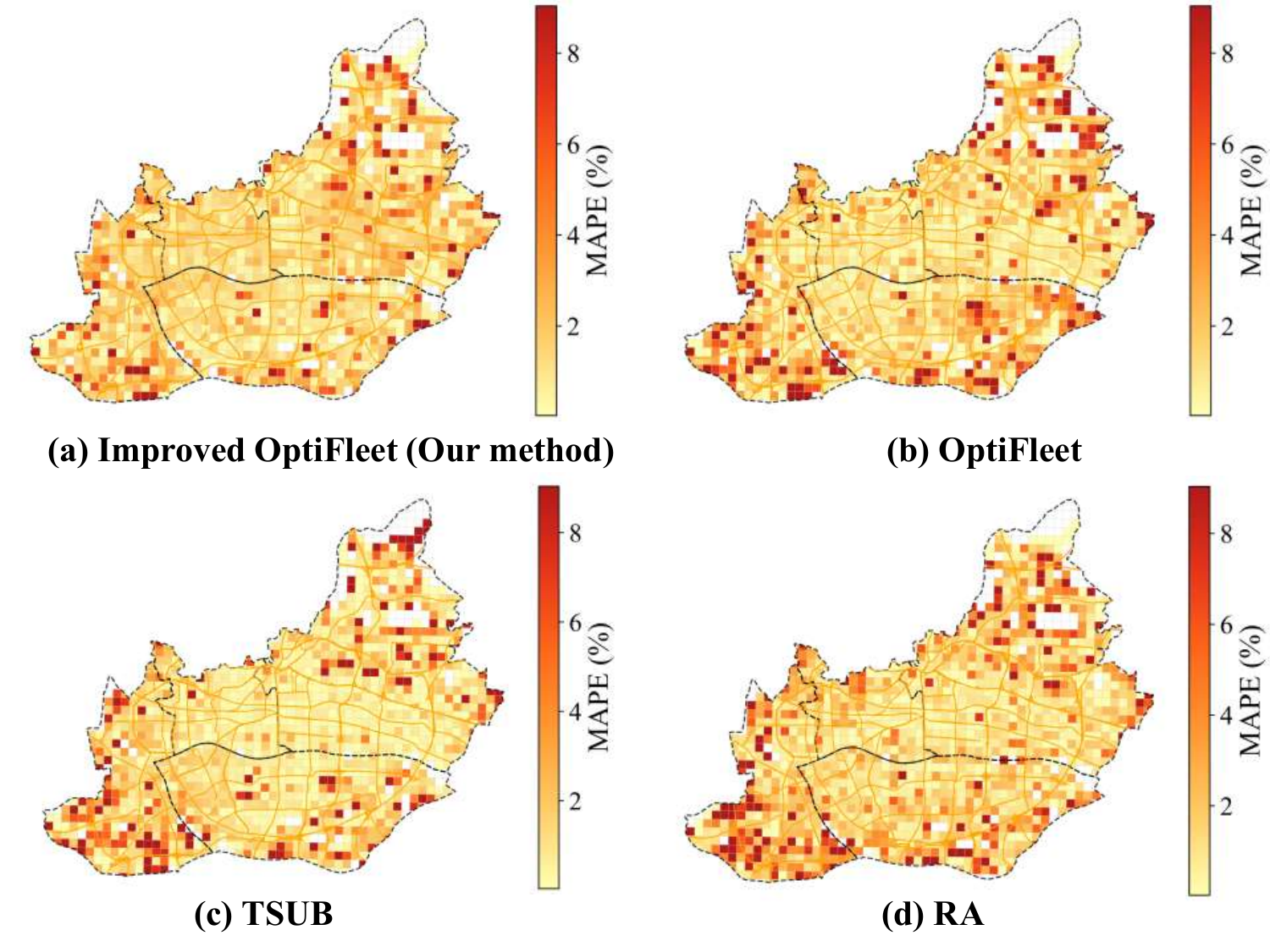}
  \caption{Spatial distribution of MAPE errors for PM\textsubscript{2.5} estimation under different models (200 vehicles).}
  \label{fig3}
\end{figure}

Furthermore, Fig.~\ref{fig3} presents the spatial distribution of MAPE errors between the 200-vehicle estimations and the ground truth across different models for PM\textsubscript{2.5}, which is recognized as the dominant on-road pollutant \cite{zhong2025predicting}. The results demonstrate that our proposed selection strategy consistently achieves the lowest estimation errors compared to other methods. In terms of performance, the models are ranked as follows: \texttt{Improved OptiFleet} $>$ \texttt{OptiFleet} $>$ TSUB $>$ RA.

\subsection{Ablation study}

\begin{figure*}[htbp]
  \centering
  \includegraphics[width=0.9\textwidth]{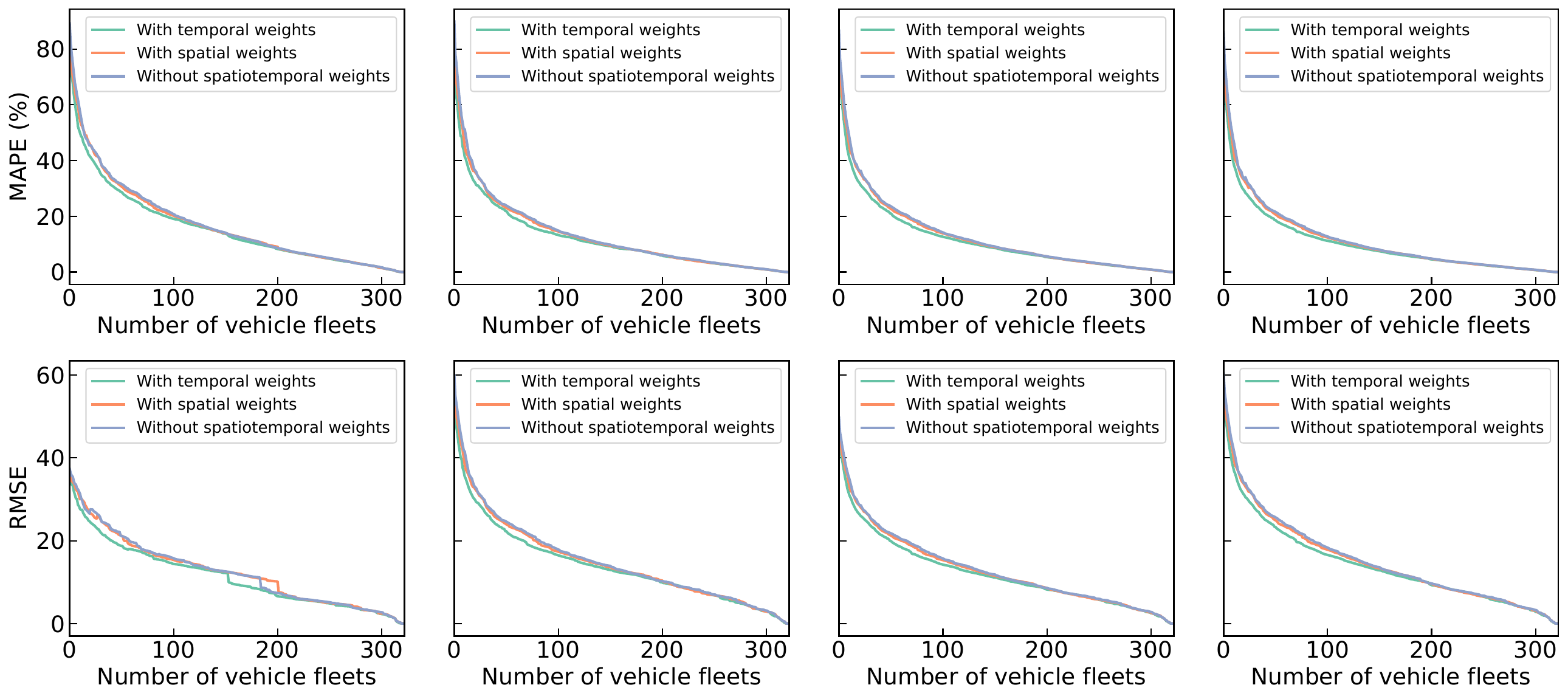}
  \caption{Ablation study on the contribution of spatial and temporal weighting to sensing utility.}
  \label{fig4}
\end{figure*}

To assess the robustness of the proposed model and examine the influence of different weighting components under the same sensing task, we conduct an ablation study focusing on the role of spatiotemporal weights in estimation accuracy. Fig.~\ref{fig4} provides a comparative analysis of model variants incorporating or omitting spatial and temporal weighting components, which are informed by heterogeneous urban data sources, including traffic indices and visual semantics extracted from street-level imagery. The ablation results highlight the critical role of temporal weighting in enhancing sensing accuracy. Temporal weighting captures dynamic urban mobility patterns, particularly variations in traffic flow, and provides essential contextual information for accurately modeling short-term fluctuations in pollutant concentrations. This finding is consistent with existing studies that emphasize the strong temporal correlation between traffic dynamics and air pollutant levels, particularly in densely populated urban environments \cite{liu2019spatial}. In comparison, spatial weighting, which is based on static street-view features, provides only marginal improvements in sensing performance. These results indicate that although spatial heterogeneity contains valuable information, it alone is insufficient to capture the real-time variability required for effective mobile sensing. These insights collectively underscore the importance of incorporating temporally dynamic urban patterns, such as congestion levels, mobility flows, and activity rhythms, in real-time sensing systems \cite{hou2025advancing, han2024exploring}. Compared to static spatial indicators, these temporal factors play a more critical role in guiding efficient sensor deployment. Prioritizing temporally informed selection strategies will be essential to enhance the effectiveness and adaptability of mobile air quality monitoring networks.

\section{Conclusions and discussion}\label{sec:conclusions}
This study presents an adaptive vehicle selection framework that integrates heterogeneous open-source urban data to enhance mobile sensing under resource constraints. 
By introducing entropy-based selection guided by spatiotemporal weights, this study proposed the \texttt{improved OptiFleet} algorithm, which effectively reduces data redundancy while improving coverage diversity. 
Experimental results demonstrate that the proposed approach achieves high estimation accuracy with a reduced fleet, maintaining MAPE below 5\% using only 200 vehicles. 
The comparison and ablation analyses reveal that temporal dynamics, especially traffic flow, play a crucial role in determining the sensing utility of mobile air quality monitoring. 
These findings emphasize the necessity of integrating dynamic urban signals into mobile sensing systems to optimize the scalability and cost-effectiveness of sensing strategies.
Nevertheless, the validation was conducted only in Guangzhou due to data availability; the generalization of the proposed framework to other cities with different traffic patterns, fleet structures, and urban morphologies remains to be explored.

\section*{Acknowledgment}
The authors would like to thank Professor Yonghong Liu from the School of Intelligent Systems Engineering, Sun Yat-sen University, for providing essential data support. Besides, this study was supported by the National Natural Science Foundation of China (Grant No.U24A20252 and No.62373315), Guangdong provincial project (Grant No.2023ZT10X009), and the Nansha Key Science and Technology Project (Grant No.2023ZD006).

\bibliographystyle{IEEEtran}
\bibliography{refs}

\end{document}